\documentclass[aps,prl,preprint,showpacs,groupedaddress]{revtex4}
\usepackage {amssymb}
\usepackage {amsmath}
\usepackage {graphicx}
\usepackage {longtable}

\begin{document}
\title{The metal-insulator transition and lattice distortion in
semiconductors}

\author{Fedor V.Prigara}
\affiliation{Institute of Physics and Technology,
Russian Academy of Sciences,\\
21 Universitetskaya, Yaroslavl 150007, Russia}
\email{fprigara@recnti.uniyar.ac.ru}

\date{\today}

\begin{abstract}

A relation between the energy of an elementary `insulating' excitation
corresponding to the metal-insulator transition and the bandgap width in a
semiconductor is obtained. An effect of atomic relaxation on the temperature
and pressure dependence of the bandgap width is considered. It is shown that
the metal-insulator transition in a semiconductor causes a weak rhombohedral
or monoclinic distortion in the case of a diamond and zincblende structure
and a weak tetragonal or orthorhombic distortion in the case of a rocksalt
structure. A change in the bandgap associated with a ferroelectric
(antiferroelectric) transition in a semiconductor is also obtained.

\end{abstract}


\maketitle

Recent experiment on the detwinning of single crystals of iron
pnictide superconductors by a weak uniaxial pressure [1] gives
evidence for the metal-insulator transition in underdoped pnictide
superconductors. Thus, iron pnictide superconductors are
superconducting narrow bandgap semiconductors, similarly to the
case of cuprate superconductors. Here we consider the
metal-insulator transition in semiconductors, including
superconducting semiconductors such as lead telluride, cuprates,
ruthenates, and iron pnictides. In most of insulators and
semiconductors, the energy $E_{i} \left( {0} \right)$ of an
elementary `insulating' excitation which determines the
metal-insulator transition temperature is equal to the bandgap
width $E_{g} \left( {0} \right)$ at zero temperature (0
\textit{K}) [2]. Here we infer a more general relation between the
energy of an elementary `insulating' excitation and the bandgap
width, which holds for such narrow bandgap semiconductors as lead
chalcogenides and indium pnictides. We show that the temperature
and pressure (in the high-pressure region) dependence of the
bandgap in a semiconductor is caused by atomic relaxation
associated with transverse optical modes. A change in the bandgap
caused by structural transitions such as ferroelectric and
antiferroelectric transitions is considered. The metal-insulator
transition causes a weak lattice distortion of a ferroelastic type
[3] in semiconductors with a diamond, zincblende, and rocksalt
structure, which produces a decrease in the coordination number
\textit{z}. This weak ferroelastic distortion leads, for example,
to a perfect cubic cleavage in single crystals of lead
chalcogenides which depends on the temperature [4] and vanishes at
the metal-insulator transition temperature $T_{MI} $ .

According to a general criterion for phase transitions in
crystalline solids [2], there is a general relation between the
energy $E_{i} \left( {0} \right)$ of an elementary `insulating'
excitation at zero temperature (0 \textit{K}) and the
metal-insulator transition temperature $T_{MI} $ of the form

\begin{equation}
\label{eq1}
E_{i} \left( {0} \right) = \alpha k_{B} T_{MI} ,
\end{equation}

\noindent
where $k_{B} $ is the Boltzmann constant, and $\alpha = 18$.

An elementary insulating excitation is a collective many-particle excitation
which invokes a local lattice distortion within a crystalline domain with a
size of $d_{c} \approx 180nm$ [2]. The energy $E_{i} \left( {0} \right)$ of
an elementary insulating excitation is proportional to the bandgap width
$E_{g} \left( {0} \right)$ at zero temperature (0\textit{K}),

\begin{equation}
\label{eq2}
E_{i} \left( {0} \right) = kE_{g} \left( {0} \right).
\end{equation}

In most of insulators and semiconductors, \textit{k}=1. A hydrogen-like
model with the (high-frequency) dielectric constant $\varepsilon $
accounting for many-body effects (the crystal structure and electron
correlations) gives an estimation of the bandgap width $E_{g} $ in the form

\begin{equation}
\label{eq3}
E_{g} = \frac{{Z^{2}}}{{\varepsilon ^{2}}}Ry,
\end{equation}

\noindent
where \textit{Z} is the mean effective charge of ions (in the units of the
electron charge \textit{e}), and $Ry = 13.6eV$ is the Rydberg constant. A
charge \textit{Z} is close to $Z = 3.5$ for diamond, silicon, and germanium,
is close to $Z = 4.0$ for III-V semiconductors (GaAs, GaSb), and is close to
$Z = 4.5$ for II-VI semiconductors (cadmium and zinc chalcogenides). In
silica ($SiO_{2} $), the dielectric constant is $\varepsilon = 3.85$, and
the bandgap width is $E_{g} \left( {0} \right) \approx 10eV$ [5], so that $Z
\approx 3.5$, as in silicon. In $CaWO_{4} $ with a tetragonal crystal
structure, the dielectric constant is $\varepsilon = 8$, and the bandgap
width is $E_{g} \left( {0} \right) \approx 4.2eV$ [6], so that $Z \approx
4.5$, as in II-VI semiconductors. In cuprates and ruthenates, a charge
\textit{Z} is close to $Z \approx 2$ ($La_{2} CuO_{4} $), or $Z \approx 2.5$
($Ca_{2} RuO_{4} $) (see below).

In narrow bandgap semiconductors with a rocksalt and zincblende structure,
the coefficient \textit{k} in the equation (\ref{eq2}) is an integer number which is
larger than 1. In lead chalcogenides, the value of the coefficient
\textit{k} is close to the coordination number $z = 6$ in the rocksalt
structure. In indium pnictides, InSb and InAs, the value of the coefficient
\textit{k} is close to the coordination number $z = 3$ corresponding to a
rhombohedral distortion of the zincblende structure.

In lead chalcogenides, the metal-insulator transition temperature $T_{MI} $
corresponds to a temperature at which their cubic cleavage vanishes [4]. The
equation (\ref{eq1}) gives the energy $E_{i} \left( {0} \right)$ of an elementary
`insulating' excitation, and the equation (\ref{eq2}) gives the value of the
coefficient \textit{k} for lead chalcogenides:

\newcommand{\PreserveBackslash}[1]{\let\temp=\\#1\let\\=\temp}
\let\PBS=\PreserveBackslash
\begin{longtable}
{|p{85pt}|p{85pt}|p{85pt}|p{85pt}|p{85pt}|}
a & a & a & a & a  \kill
\hline
Semiconductor&
\[
E_{g} \left( {0} \right)\left( {eV} \right)
\]
&
\[
T_{MI} \left( {K} \right)
\]
&
\[
E_{i} \left( {0} \right)\left( {eV} \right)
\]
&
\[
k
\]
 \\
\hline
\textbf{}PbS&
0.29&
935&
1.45&
5 \\
\hline
\textbf{}PbSe&
0.165&
640&
0.99&
6 \\
\hline
\textbf{}PbTe&
0.19&
610&
0.95&
5 \\
\hline
\end{longtable}

\textbf{}

The temperature dependence of the bandgap width $E_{g} \left( {T} \right)$
and the high-frequency dielectric constant $\varepsilon \left( {T} \right)$
in lead chalcogenides follows the relation (\ref{eq3}) with Z=const [4].

The value of the photo-voltage in lead chalcogenide films (times
the charge of an electron \textit{e}) is determined by the energy
$E_{i} $ of an elementary insulating excitation and is much higher
than the bandgap width $E_{g} $ [4]. A non-radiative recombination
in a semiconductor is connected with the creation of an elementary
insulating excitation with a subsequent heating of the lattice in
the process of the decay of this excitation. In the case of lead
chalcogenides, the energy of an elementary insulating excitation
is accumulated by the Auger recombination [4].

The energy $E_{i} \left( {T} \right)$ of an elementary insulating excitation
depends on the temperature \textit{T} as follows [2]

\begin{equation}
\label{eq4}
E_{i} \left( {T} \right) = E_{i} \left( {0} \right) - \beta k_{B} T.
\end{equation}

A dimensionless atomic relaxation constant $\beta $ is normally close to 6.
Its value is determined by the excitation of two transverse optical modes
along the three crystallographic axes with a maximum value of the
propagation vector $q = q_{max} $. A decrease in the value of $\beta $ to
$\beta = 5$ in the case of cadmium chalcogenides is caused by the
suppression of one of these atomic relaxation modes and can be attributed to
a monoclinic lattice distortion associated with the metal-insulator
transition. Similar is the case of silicon and germanium. In the case of
zinc chalcogenides, the value of $\beta $ is $\beta = 6$. This value
corresponds to a rhombohedral or the c-axis distortion for a zincblende and
hexagonal structure, respectively.

The equation (\ref{eq4}) determines the temperature dependence $E_{g} \left( {T}
\right)$ of the bandgap width for most of semiconductors (with $E_{g} =
E_{i} $). For indium antimonide InSb, $k = 3$, and the equations (\ref{eq2}) and (\ref{eq4})
give

\begin{equation}
\label{eq5}
E_{i} = 3E_{g} = E_{i} \left( {0} \right) - \beta k_{B} T,
\end{equation}

\noindent
so that the temperature dependence of the bandgap width in InSb is given by
the equation

\begin{equation}
\label{eq6}
E_{g} \left( {T} \right) = E_{g} \left( {0} \right) - \frac{{1}}{{3}}\beta
k_{B} T \approx E_{g} \left( {0} \right) - 2k_{B} T,
\end{equation}

\noindent
where $\beta \approx 6$.

The temperature dependence of the bandgap width in lead chalcogenides is
anomalous and at low temperatures below $T_{0} $, where $E_{g} \left( {T}
\right)$ has a maximum, has the form

\begin{equation}
\label{eq7}
E_{g} \left( {T} \right) = E_{g} \left( {0} \right) + \beta k_{B} T,
\end{equation}

\noindent
where $\beta \approx 5$.

The temperature $T_{0} $ is determined by the equation

\begin{equation}
\label{eq8}
E_{i} \left( {T_{0}}  \right) = E_{i} \left( {0} \right) - \beta k_{B} T_{0}
= z_{0} E_{g} \left( {T_{0}}  \right) = 2\left( {E_{g} \left( {0} \right) +
\beta k_{B} T_{0}}  \right),
\end{equation}

\noindent
where $z_{0} = 2$ is a coordination number in an orthorhombically distorted
rocksalt structure. Above $T_{0} $, the energy $E_{i} \left( {T} \right)$ of
an elementary insulating excitation in lead chalcogenides is proportional to
the bandgap width $E_{g} \left( {T} \right)$,

\begin{equation}
\label{eq9}
E_{i} \left( {T} \right) = z_{0} E_{g} \left( {T} \right) = 2E_{g} \left(
{T} \right).
\end{equation}

The $E_{i} \left( {T} \right)$dependence is given by the equation (\ref{eq4}). The
value of the temperature $T_{0} $ at which $E_{g} \left( {T} \right)$ has a
maximum is close to $T_{0} \approx 440K$ for PbTe, $T_{0} \approx 510K$ for
PbSe, and $T_{0} \approx 680K$ for PbS [4].

An atomic relaxation constant $\beta $ determines also a change in the
bandgap width caused by a structural transition of a ferroelectric
(antiferroelectric) type with the transition temperature (the Curie
temperature) $T_{s} $, in accordance with the equation

\begin{equation}
\label{eq10}
E_{g} \left( {0} \right) - E_{i} \left( {0} \right) = \beta k_{B} T_{s} .
\end{equation}

Bismuth ferrite ($BiFeO_{3} $) exhibits a metal-insulator transition at
$T_{MI} \approx 1200K$ and a ferroelectric transition at $T_{s} \approx
1100K$ [7]. The equation (\ref{eq1}) gives $E_{i} \left( {0} \right) = 1.85eV$ ($k =
1$) and the equation (\ref{eq10}) with $\beta \approx 6$ gives $E_{g} \left( {0}
\right) \approx 2.4eV$. According to the equation (\ref{eq4}) with $\beta \approx
6$, the bandgap width $E_{g} \left( {T_{MI}}  \right)$ just below the
metal-insulator transition temperature $T_{MI} $ is equal to $E_{g} \left(
{T_{MI}}  \right) \approx 1.3eV$. Optical absorption measurements give the
values $E_{g} \left( {0} \right) \approx 2.5eV$ and $E_{g} \left( {T_{MI}}
\right) \approx 1.5eV$ [7].

Another perovskite compound, barium bismuthate ($BaBiO_{3} $) undergoes a
metal-insulator transition at $T_{MI} = 893K$ and a structural transition of
an antiferroelectric type at $T_{s} = 425K$ [8]. The $\alpha $ phase (below
$T_{s} $) is monoclinic, the $\beta $ phase (between $T_{s} $ and $T_{MI} $)
is rhombohedral, and the $\gamma $ phase (above $T_{MI} $) is cubic.
According to equations (\ref{eq1}) and (\ref{eq10}), $E_{i} \left( {0} \right) \approx
1.4eV$ and $E_{g} \left( {0} \right) \approx 1.6eV$.

Barium bismuthate doped with K, $Ba_{1 - x} K_{x} BiO_{3} $, is a
high-temperature superconductor of an improper (charge-ordered)
type [1] with the maximum superconducting transition temperature
$T_{c} = 30K$ at $x = 0.25$ [9]. Superconductivity in $Ba_{1 - x}
K_{x} BiO_{3} $ occurs only in a cubic metallic phase [9]. The
corresponding charge gap in the superconducting phase is about
$\Delta _{ch} \approx 0.05eV$ (see ref. 1). There are other
superconducting semiconductors, such as $SrTiO_{3} $ (doped with
Nb) [10] and PbTe (doped with Tl) [11]. In the last case, the
thallium doping seems to be equivalent to the doping with an
additional non-stoichiometric Te, so that a composition $Pb_{1 -
x} Tl_{x} Te$ corresponds to a composition $PbTe_{1 + x/2} $. The
optimal doping level with a highest $T_{c} $ has not been achieved
due to a low solubility limit of Tl ($x_{0} = 0.015$). $PbTe_{1 +
\delta}  $ should be a superconductor analogous to $FeSe_{1 -
\delta}  $ [12].

Cuprates and ruthenates are also superconducting semiconductors.
For example, $La_{2} CuO_{4} $ exhibits a metal-insulator
transition (coinciding with an antiferromagnetic transition) at
$T_{MI} = T_{N} = 240K$. The equation (\ref{eq1}) gives the
bandgap width of $E_{g} \left( {0} \right) = 0.37eV$. $La_{2}
CuO_{4.093} $ is a high-temperature superconductor of an improper
(charge-ordered) type with the superconducting transition
temperature $T_{c} = 45K$ [13]. $Ca_{2} RuO_{4 + \delta}  $
undergoes a metal-insulator transition at $T_{MI} = 357K$ for
$\delta = 0$ [14]. According to the equation (\ref{eq1}), the
bandgap width is $E_{g} \left( {0} \right) = 0.55eV$. The
metal-insulator transition temperature $T_{MI} $ and the bandgap
width $E_{g} \left( {0} \right)$ decrease with increasing $\delta
$ [14] similarly to the case of cuprates. There should be the
superconducting phase for sufficiently high $\delta $. Similar is
the case of $Sr_{2} RuO_{4 + \delta } $. Due to a larger atomic
radius of Sr as compared with those of Ca, the non-stoichiometric
oxygen is presumably present on interstitial sites in the
superconducting $Sr_{2} RuO_{4} $. In $\left( {Ca_{1 - x} Sr_{x}}
\right)_{2} RuO_{4} $, the metal-insulator transition temperature
$T_{MI} \left( {x} \right)$ goes to zero at $x_{0} \approx 0.08$
[14]. There should be the superconducting phase in the region of
doping $x \approx 0.01 - 0.02$. A decrease in the magnetic
susceptibility $\chi \left( {T} \right)$ of $\left( {Ca_{1 - x}
Sr_{x}}  \right)_{2} RuO_{4} $ at low temperatures in the doping
range $x \approx 0.01 - 0.02$ is caused by the presence of the
superconducting phase with $T_{c} \approx 10K$ for $x = 0.01$.
Above $T_{c} $, the Curie law is valid [14],

\begin{equation}
\label{eq11}
\chi \left( {T} \right) \propto T^{ - 1}.
\end{equation}

Recent experiment on the detwinning of single crystals of iron
pnictide superconductors by a weak uniaxial pressure [1] shows
that the samples have a multi-phase composition in the underdoped
region. The Co doping stabilizes the $Ba\left(Fe_{1-x}
Co_{x}\right)_{2} As_{2} $ phases (products of the partial
decomposition of $Ba Fe_{2} As_{2} $ should also be present).
However, two phases coexist in the doping range between $x=0.025 $
and $x=0.061 $, an insulating antiferromagnetic phase with a low
\textit{x}, and a metallic superconducting phase with a high
\textit{x}. These two phases form an oriented
quasi-two-dimensional epitaxial structure [15], which can explain
the a-b anisotropy of the in-plane resistivity. The thickness of
the monophase layers has an order of the radius of the atomic
relaxation region, i.e. has an order of 8 nm [16].

Both the metal-insulator transition temperature and the Neel
temperature decrease with increasing doping level \textit{x},
similarly to the case of $\left( {Ca_{1 - x} Sr_{x}}  \right)_{2}
RuO_{4} $ [14]. Iron pnictide superconductors are superconducting
semiconductors similarly to cuprate and ruthenate superconductors
.

A weak uniaxial pressure in this experiment orients the
quasi-two-dimensional epitaxial structure formed by the two phases
of $Ba\left(Fe_{1-x} Co_{x}\right)_{2} As_{2} $ and changes
relative volumes of the insulating and metallic phases (a partial
superconducting transition for $x=0.016 $ and $x=0.025 $) [1].

The pressure dependence of the energy \textit{E} of an elementary
excitation in the high-pressure region is given by the equation
[16]

\begin{equation}
\label{eq12} E\left( {P} \right) = E\left( {0} \right) - \alpha
_{P} P/n_{0} ,
\end{equation}

\noindent
where $n_{0} \approx 1.1 \times 10^{22}cm^{ - 3}$ is a constant, \textit{P}
is the pressure, and $\alpha _{P} $ is the atomic relaxation constant,
$\alpha _{P} = 2$ for ferroelectric, antiferroelectric, and metal-insulator
transitions.

This value of $\alpha _{P} $ is determined by the excitation of two
transverse optical modes with $q = q_{max} $ along the polarization axis in
the case of ferroelectric and antiferroelectric transitions and along the
easy ferroelastic axis in the case of metal-insulator transitions (the axis
along which a main lattice distortion associated with the metal-insulator
transition is directed).

The value of the atomic relaxation constant $\alpha _{P} $ for ferromagnetic
transitions is $\alpha _{P} = 4$ and corresponds to transverse optical modes
with $q = q_{max} $ in the plane perpendicular to the magnetization vector.

The value of the atomic relaxation constant for the melting transition is
$\alpha _{P} = 18$ and can be attributed to the excitation of three
acoustical and three optical modes along the three crystallographic axes.

A weak ferroelastic distortion associated with the metal-insulator
transition in semiconductors determines the shape of single crystals. Single
crystals of diamond have an octahedral morphology due to a weak rhombohedral
distortion associated with a high-temperature metal-insulator transition.
Single crystals of lead chalcogenides have a simple cubic morphology due to
a weak tetragonal (orthorhombic) distortion caused by the metal-insulator
transition.

Similarly, a low-temperature ferroelastic transition with the
transition temperature $T_{f} \approx \theta _{D} /\alpha $, where
$\theta_{D}$ is the Debye temperature, [17] produces a weak
rhombohedral distortion in fcc metals, a weak tetragonal (or
orthorhombic) distortion in bcc metals, and a weak c-axis
distortion in hcp metals. In insulators, a low-temperature
ferroelastic distortion corresponds to a relative expansuion of
the lattice at zero temperature and produces a negative thermal
expansion below the ferroelastic transition temperature $T_{f} $.
For example, in $MnF_{2} $ a low-temperature ferroelastic
transition occurs at $T_{f} \approx 20K$ and is marked by a
minimum in the temperature dependence of the unit cell volume
[18]. The amplitude $\delta = \Delta a/a$ (\textit{a} is the
lattice parameter) of a low-temperature ferroelastic distortion in
$MnF_{2} $ is about $\delta \approx 0.5 \times 10^{ - 4}$.

A relative expansion of the lattice in insulators with decreasing
temperature below $T_{f} $ is caused by interatomic repulsion at
short interatomic distances (the Born-Mayer equation for the
cohesive energy of ionic crystals [19]).

In metals, a low-temperature ferroelastic distortion corresponds
to a relative contraction of the lattice at zero temperature and
produces a positive thermal expansion below $T_{f} $. A
ferroelastic distortion associated with the superconducting
transition [17] has an opposite sign and compensates this effect,
so that an overall thermal expansion, for example, in $MgB_{2} $
is very small below the superconducting transition temperature
$T_{c} $ [20].

To summerize, we show that there is a general relation between the
energy of an elementary insulating excitation and the
metal-insulator transition temperature which is valid for narrow
bandgap semiconductors too. We give an estimation of the bandgap
width in insulators and semiconductors. The temperature and
pressure (in the high-pressure region) dependence of of the
bandgap width is obtained. We show that there is a weak lattice
distortion of a ferroelastic type associated with the
metal-insulator transition in semiconductors.

\begin{center}
---------------------------------------------------------------
\end{center}

[1] J.-H.Chu, J.G.Analytis, K.De Greve, P.L.McMahon, Z.Islam,
Y.Yamamoto, and I.R.Fisher, arXiv:1002.3364 (2010).

[2] F.V.Prigara, arXiv:0902.4350 (2009).

[3] S.H.Curnoe and A.E.Jacobs, Phys. Rev. B \textbf{64}, 064101 (2001).

[4] Yu.I.Ravich, B.A.Efimova, and I.A.Smirnov, \textit{Semiconducting Lead
Chalcogenides} (Plenum Press, New York, 1970).

[5] V.A.Gritsenko, Usp. Fiz. Nauk \textbf{179}, 921 (2009) [Physics-Uspekhi
\textbf{52}, 869 (2009)].

[6] N.G.Ryabtsev, \textit{Materials for Quantum Electronics} (Soviet Radio
Press, Moscow, 1972).

[7] R.Palai, R.S.Katiyar, H.Schmid et al., Phys. Rev. B \textbf{77}, 014110
(2008).

[8] Q.Zhou and B.J.Kennedy, Solid State Commun. \textbf{132}, 389 (2004).

[9] D.G.Hinks, B.Dabrowski, J.D.Jorgensen, A.W.Mitchell, D.R.Richards,
S.Pei, and D.Shi, Nature \textbf{333}, 836 (1988).

[10] A.Bussmann-Holder, A.R.Bishop, and A.Simon, arXiv:0909.4176 (2009).

[11] Y.Matsushita, P.A.Wianecki, A.T.Sommer, T.H.Geballe, and
I.R.Fisher, Phys. Rev. B \textbf{74}, 134512 (2006).

[12] F.C.Hsu, J.Y.Luo, K.W.Yeh et al., Proc. Nat. Aca. Sci. USA
\textbf{105}, 14262 (2008).

[13] T.Hirayama, M.Nakagawa, and Y.Oda, Solid State Commun. \textbf{113},
121 (1999).

[14] S.G.Ovchinnikov, Usp. Fiz. Nauk \textbf{173}, 27 (2003)
[Physics-Uspekhi \textbf{46}, 21 (2003)].

[15] P.K.Mang, S.Larochelle, A.Mehta et al., Phys. Rev. B
\textbf{70}, 094507 (2004).

[16] F.V.Prigara, arXiv:0708.1230 (2007).

[17] F.V.Prigara, arXiv:1001.4152 (2010).

[18] R.Schleck, Y.Nahas, R.P.S.M.Lobo, J.Varignon, M.B.Lepetit,
C.S.Nelson, and R.L.Moreira, arXiv:0910.3137 (2009).

[19] V.S.Urusov, \textit{Theoretical Crystal Chemistry} (Moscow
University Press, Moscow, 1987).

[20] J.J.Neumeier, T.Tomita, M.Debessai, J.S.Schilling,
P.W.Barnes, D.G.Hinks, and J.D.Jorgensen, Phys. Rev. B
\textbf{72}, 220505(R) (2005).

\end{document}